\documentclass[11pt, logo, nonumbering]{preprint}
\usepackage[utf8]{inputenc}
\usepackage{microtype}
\usepackage{graphicx}
\usepackage{amsmath}
\usepackage{booktabs}
\usepackage{xcolor}
\usepackage{multirow}
\usepackage{algorithm}
\usepackage{algpseudocode}
\usepackage{listings}
\usepackage{tikz}
\usetikzlibrary{shapes.geometric, arrows.meta, positioning, fit, backgrounds}
\definecolor{mydarkblue}{rgb}{0,0.08,0.45}
\usepackage[colorlinks=true,linkcolor=mydarkblue,citecolor=mydarkblue,filecolor=mydarkblue,urlcolor=mydarkblue]{hyperref}
\usepackage{xspace}
\usepackage{cleveref}

\usepackage[style=numeric-comp,maxbibnames=3,minnames=1,backend=bibtex, doi=false,eprint=false,url=false]{biblatex}

\definecolor{gred}{RGB}{250, 210, 207}
\definecolor{coolblue1}{rgb}{0.91, 0.94, 0.98}
\definecolor{coolblue2}{rgb}{0.76, 0.85, 0.94}
\definecolor{coolblue3}{rgb}{0.54, 0.72, 0.87}
\definecolor{coolblue4}{rgb}{1, 1, 1}
\definecolor{cachegreen}{RGB}{200, 235, 200}
\definecolor{missred}{RGB}{255, 220, 220}

\newcommand{\sysname}{\textsc{VoiceAgentRAG}\xspace}
\newcommand{\slowthinker}{\textit{Slow Thinker}\xspace}
\newcommand{\fasttalker}{\textit{Fast Talker}\xspace}

\begin{document}

\title{\sysname: Solving the RAG Latency Bottleneck in Real-Time Voice Agents Using Dual-Agent Architectures}

\author{
Jielin Qiu, Jianguo Zhang, Zixiang Chen, Liangwei Yang, Ming Zhu, Juntao Tan,  \\
Haolin Chen, Wenting Zhao, Rithesh Murthy, Roshan Ram, Akshara Prabhakar, \\
Shelby Heinecke, Caiming, Xiong, Silvio Savarese, Huan Wang \\
~~~~\\
\textsuperscript{}Salesforce AI Research \\
~~~~\\
\href{https://github.com/SalesforceAIResearch/VoiceAgentRAG}{
  \raisebox{-0.3\height}{\includegraphics[height=0.8cm]{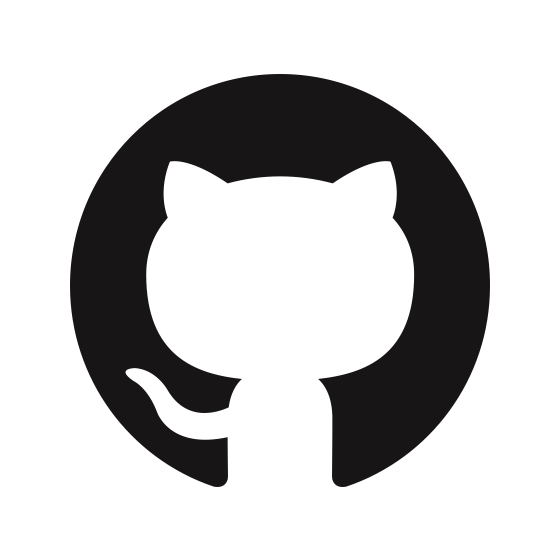}}
  \textbf{https://github.com/SalesforceAIResearch/VoiceAgentRAG}
}
}

\maketitle

\begin{abstract}
Retrieval-Augmented Generation (RAG) enables language models to ground their responses in external knowledge, but the retrieval step introduces latency that is incompatible with real-time voice conversations. A typical vector database query adds 50--300ms to the response pipeline, which, combined with embedding computation and LLM generation, pushes total latency well beyond the 200ms budget required for natural conversational flow. We present \sysname, an open-source \textbf{dual-agent memory router} that decouples retrieval from response generation. A background \slowthinker agent continuously monitors the conversation stream, predicts likely follow-up topics using an LLM, and pre-fetches relevant document chunks into a FAISS-backed semantic cache. A foreground \fasttalker agent reads only from this sub-millisecond cache, bypassing the vector database entirely on cache hits. We evaluate \sysname on a 200-query benchmark across 10 diverse conversation scenarios using Qdrant Cloud as the production vector database, achieving a \textbf{75\% overall cache hit rate} (79\% on warm turns) and a \textbf{316$\times$ retrieval speedup} (110ms $\to$ 0.35ms) on cache-hit queries, saving 16.5 seconds of cumulative retrieval latency across 150 cache hits. 
\end{abstract}

\section{Introduction}
\label{sec:intro}

The emergence of voice-enabled AI agents has created a new class of real-time systems where latency is a first-class design constraint. Unlike text-based chatbots where users tolerate multi-second response times, voice conversations require sub-200ms response latency to feel natural~\cite{ethiraj2025telecom}. This budget must accommodate the entire pipeline: speech-to-text (STT), context retrieval, LLM generation, and text-to-speech (TTS).

Retrieval-Augmented Generation (RAG)~\cite{lewis2020rag} has become the standard approach for grounding LLM responses in domain-specific knowledge. However, the retrieval step introduces a critical latency bottleneck in voice pipelines. A production vector database query (e.g., to Qdrant~\cite{qdrant2024}, Pinecone, or Weaviate) typically incurs 50--300ms of network round-trip latency, which alone can consume the entire latency budget.

Existing approaches to this problem fall into three categories, none of which fully address the voice latency constraint:

\begin{itemize}
    \item \textbf{Faster retrieval infrastructure.} Optimized vector indices (FAISS~\cite{johnson2019faiss}, HNSW \cite{Malkov2016EfficientAR}) reduce search time but do not eliminate network latency when the database is hosted remotely, which is the common production deployment pattern.
    \item \textbf{Semantic caching.} Systems like GPTCache~\cite{gptcache2023}, QVCache~\cite{gocer2025qvcache}, and the Semantic Lookaside Buffer~\cite{arslan2025aeon} cache previous query results, but they are reactive: they only help when the \textit{exact same} or very similar query recurs.
    \item \textbf{Speculative retrieval.} Stream RAG~\cite{arora2025streamrag} and Speculative RAG~\cite{wang2024specrag} predict what to retrieve during generation, but they operate within a single query's lifecycle rather than across conversation turns.
\end{itemize}

We propose \sysname, a dual-agent architecture inspired by Kahneman's System 1/System 2 framework~\cite{kahneman2011thinking} and recent work on dual-process AI agents~\cite{zhang2025dptagent, du2025cdr}. The key insight is that in multi-turn voice conversations, \textit{the next question is often predictable from the current conversation context}. A customer asking about pricing will likely follow up about the enterprise plan, discounts, or billing---topics that can be pre-fetched \textit{during the current response} before the user even speaks.

\sysname consists of two concurrent agents:

\begin{enumerate}
    \item The \textbf{\slowthinker} runs as a background async task. On each user utterance, it uses an LLM to predict 3--5 likely follow-up topics, retrieves relevant document chunks from the vector database, and pre-populates a FAISS-backed semantic cache. This work happens in the background while the user is listening to the current response.

    \item The \textbf{\fasttalker} handles user queries in the foreground. It first checks the semantic cache (sub-millisecond lookup), and only falls back to the vector database on cache miss. When it does fall back, it caches the retrieved results for future queries.
\end{enumerate}

Our contributions are:
\begin{itemize}
    \item An open-source dual-agent memory router for voice agents that decouples retrieval from response generation.
    \item A document-embedding-indexed semantic cache that achieves sub-millisecond lookup with correct semantic matching.
    \item A comprehensive 200-query evaluation across 10 conversation scenarios on a production Qdrant Cloud deployment, demonstrating 75\% cache hit rate and 316$\times$ retrieval speedup.
    \item Analysis of cache behavior across conversation depth, topic patterns, and cold/warm start conditions.
\end{itemize}

\section{Method}
\label{sec:method}

\subsection{Architecture Overview}

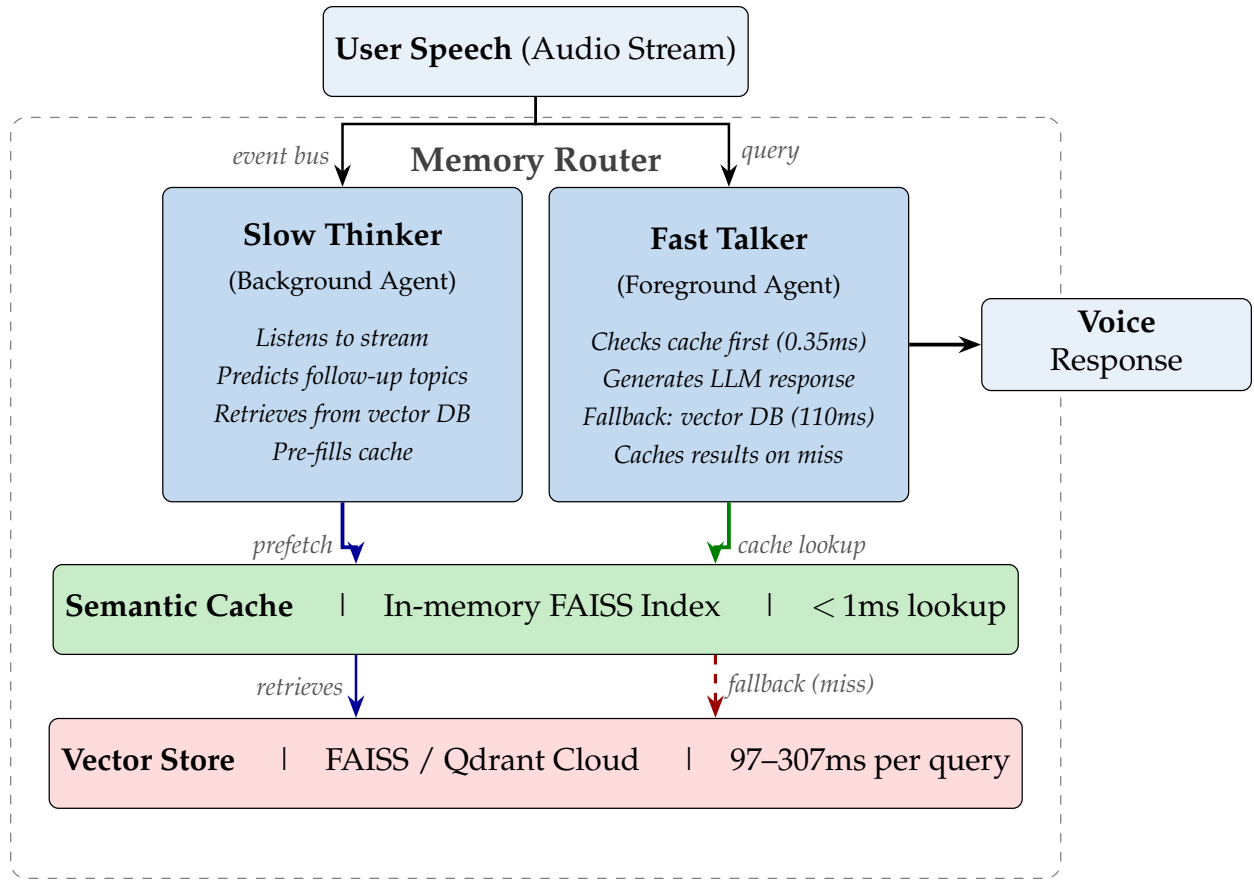
\begin{figure}[htp]
    \centering
    \resizebox{\linewidth}{!}{%
    \begin{tikzpicture}[
        node distance=0.8cm and 1.2cm,
        box/.style={rectangle, draw, rounded corners=3pt, minimum height=1.0cm, minimum width=3.0cm, align=center, font=\small},
        agent/.style={box, minimum width=4.0cm, minimum height=3.5cm},
        cache/.style={box, fill=cachegreen, minimum width=9.5cm, minimum height=1.0cm},
        store/.style={box, fill=missred, minimum width=9.5cm, minimum height=1.0cm},
        arr/.style={-{Stealth[length=2.5mm]}, thick},
        lbl/.style={font=\scriptsize\itshape, text=gray!70!black},
    ]
    \node[box, fill=coolblue1] (user) {\textbf{User Speech} (Audio Stream)};

    \node[agent, fill=coolblue2, below=1.0cm of user, xshift=-2.15cm] (slow) {
        \textbf{Slow Thinker}\\[2pt]
        {\scriptsize (Background Agent)}\\[6pt]
        {\scriptsize\itshape Listens to stream}\\
        {\scriptsize\itshape Predicts follow-up topics}\\
        {\scriptsize\itshape Retrieves from vector DB}\\
        {\scriptsize\itshape Pre-fills cache}
    };

    \node[agent, fill=coolblue2, below=1.0cm of user, xshift=2.15cm] (fast) {
        \textbf{Fast Talker}\\[2pt]
        {\scriptsize (Foreground Agent)}\\[6pt]
        {\scriptsize\itshape Checks cache first (0.35ms)}\\
        {\scriptsize\itshape Generates LLM response}\\
        {\scriptsize\itshape Fallback: vector DB (110ms)}\\
        {\scriptsize\itshape Caches results on miss}
    };

    \node[cache, below=5.2cm of user] (cache) {\textbf{Semantic Cache} \quad\textbar\quad In-memory FAISS Index \quad\textbar\quad $<$\,1ms lookup};

    \node[store, below=0.7cm of cache] (vs) {\textbf{Vector Store} \quad\textbar\quad FAISS / Qdrant Cloud \quad\textbar\quad 97--307ms per query};

    \node[box, fill=coolblue1, right=0.8cm of fast] (output) {\textbf{Voice}\\Response};

    \begin{scope}[on background layer]
        \node[draw=gray, dashed, rounded corners=5pt, fill=white,
              fit=(slow)(fast)(cache)(vs),
              inner sep=12pt, inner ysep=22pt] (router) {};
    \end{scope}
    \node[font=\normalsize\bfseries, text=gray!50!black,
          anchor=north] at ([yshift=-6pt]router.north) {Memory Router};

    \draw[arr] (user.south) -- ++(0,-0.3) -| node[near end, left, lbl] {event bus} (slow.north);
    \draw[arr] (user.south) -- ++(0,-0.3) -| node[near end, right, lbl] {query} (fast.north);

    \draw[arr, blue!60!black, line width=1.2pt]
        (slow.south) -- ++(0,-0.5) -| node[pos=0.15, left, lbl] {prefetch} ([xshift=-2.0cm]cache.north);
    \draw[arr, green!50!black, line width=1.2pt]
        (fast.south) -- ++(0,-0.5) -| node[pos=0.15, right, lbl] {cache lookup} ([xshift=2.0cm]cache.north);

    \draw[arr, blue!60!black]
        ([xshift=-2.0cm]cache.south) -- node[left, lbl] {retrieves} ([xshift=-2.0cm]vs.north);
    \draw[arr, dashed, red!60!black, line width=1.0pt]
        ([xshift=2.0cm]cache.south) -- node[right, lbl] {fallback (miss)} ([xshift=2.0cm]vs.north);

    \draw[arr, line width=1.2pt] (fast.east) -- (output.west);

    \end{tikzpicture}%
    }
    \caption{Architecture of \sysname. The \slowthinker (left) continuously monitors the conversation stream, predicts follow-up topics, retrieves from the vector store, and populates the semantic cache. The \fasttalker (right) checks the cache first ($<$1ms), bypassing the vector store on hits. On misses, it falls back to direct retrieval (dashed red) and caches the results for future queries.}
    \label{fig:architecture}
\end{figure}

\sysname is organized around five core components, shown in \Cref{fig:architecture}: (1) a \textbf{Memory Router} that orchestrates the system, (2) a \textbf{Conversation Stream} that provides an async event bus, (3) a \textbf{\slowthinker} background agent, (4) a \textbf{\fasttalker} foreground agent, and (5) a \textbf{Semantic Cache} backed by an in-memory FAISS index.

The data flow proceeds as follows. When a user speaks, the Memory Router publishes a \texttt{UserUtterance} event to the Conversation Stream. Both agents receive this event concurrently:

\begin{itemize}
    \item The \fasttalker immediately embeds the query, checks the semantic cache, and either serves cached context (hit) or falls back to the vector store (miss). It then generates a streaming LLM response.
    \item The \slowthinker (running as a background \texttt{asyncio} task) processes the same utterance: it retrieves context for the current query, predicts follow-up topics via the LLM, and pre-fetches document chunks for those predictions into the cache.
\end{itemize}

Because the \slowthinker operates asynchronously, its prefetching overlaps with the current turn's response generation and the user's listening/thinking time. By the time the user asks their next question, the relevant context is already in the cache.

\subsection{Semantic Cache}

The semantic cache is the critical bridge between the two agents. It is implemented as an in-memory FAISS \texttt{IndexFlatIP} (inner product on L2-normalized vectors, equivalent to cosine similarity) that stores document chunks indexed by their \textit{own} embeddings.

\textbf{Why document embeddings?} An early design indexed cached entries by the \textit{prediction query's embedding}. This produced cache ``hits'' that returned irrelevant chunks, where the prediction ``What are the pricing details?'' would cache pricing chunks, but when the user asked ``Is there a discount for annual billing?'', the cache would return general pricing info instead of the discount-specific content. By indexing with document embeddings, the cache performs a proper semantic search over the cached document chunks, returning the most relevant ones for the actual user query.

The cache supports the following operations:
\begin{itemize}
    \item \textbf{Put}: Store a document chunk with its embedding, relevance score, metadata, and TTL. Near-duplicates (cosine similarity $> 0.95$) are detected and updated rather than added.
    \item \textbf{Get}: Given a query embedding, return the top-$k$ cached chunks with cosine similarity above a threshold $\tau$. This operation takes $< 1$ms on a cache of 100+ entries.
    \item \textbf{Eviction}: Entries expire based on TTL (default: 300s). When the cache reaches maximum size, the least recently accessed entry is evicted (LRU policy).
\end{itemize}

\textbf{Threshold tuning.} The similarity threshold $\tau$ is a critical parameter. We found that query-to-document cosine similarity (using OpenAI \texttt{text-embedding-3-small}) typically ranges from 0.30 to 0.55, significantly lower than query-to-query similarity. Through empirical analysis (\Cref{sec:threshold}), we set $\tau = 0.40$ as the default, which balances precision (avoiding irrelevant context) with recall (capturing useful matches).

\subsection{Slow Thinker: Predictive Prefetching}

The \slowthinker subscribes to the Conversation Stream and processes events as follows:

\textbf{On \texttt{UserUtterance}:}
\begin{enumerate}
    \item \textbf{Direct retrieval}: Embed the current utterance and retrieve the top-$k$ document chunks from the vector store. Cache each chunk with its own embedding.
    \item \textbf{Prediction}: Use the LLM to predict the $n$ most likely follow-up topics based on the conversation history (sliding window of last 6 turns). The prompt instructs the LLM to generate document-style descriptions rather than questions, producing embeddings closer to actual document content.
    \item \textbf{Prefetch}: For each prediction, embed it, search the vector store, and cache the results. These operations run in parallel using \texttt{asyncio.gather}.
\end{enumerate}

\textbf{On \texttt{PriorityRetrieval}} (triggered by \fasttalker cache miss): Perform an immediate retrieval with expanded top-$k$ (2$\times$ default) to quickly populate the cache around the missed topic.

\textbf{Rate limiting}: To avoid overwhelming the vector store and embedding API, the \slowthinker enforces a minimum interval between predictions (default: 0.5s).

\subsection{Fast Talker: Cache-First Response}

The \fasttalker handles the critical latency path. For each user query:

\begin{enumerate}
    \item \textbf{Embed} the query using the embedding API ($\sim$200ms for remote APIs).
    \item \textbf{Cache lookup}: Search the semantic cache for the top-$k$ chunks with similarity $\geq \tau$ ($< 1$ms).
    \item \textbf{If cache hit}: Format the cached chunks as context and generate the LLM response. \textit{Skip the vector store entirely.}
    \item \textbf{If cache miss}: Fall back to direct vector store search, generate the response, and \textit{cache the retrieved results} so that the next similar query hits the cache. Also signal the \slowthinker via a \texttt{PriorityRetrieval} event.
\end{enumerate}

The cache-on-miss behavior is crucial: it means that the first query on any topic primes the cache, and all subsequent queries on that topic can be served from cache. Combined with the \slowthinker's predictive prefetching, this creates a compounding effect where the cache warms up rapidly over the first few turns.

\subsection{Conversation Stream}

The Conversation Stream is an async event bus implemented with \texttt{asyncio.Queue}. It supports five event types: \texttt{UserUtterance}, \texttt{AgentResponse}, \texttt{SilenceDetected}, \texttt{TopicShift}, and \texttt{PriorityRetrieval}. A sliding window maintains the last $N$ conversation turns (default: 10) for the \slowthinker's prediction context.

\section{Experimental Setup}
\label{sec:setup}

\subsection{Knowledge Base}

We construct a synthetic enterprise knowledge base for ``NovaCRM,'' a fictional AI-powered CRM platform. The knowledge base consists of 12 documents covering: company overview, product features, pricing plans, API documentation (overview, contacts endpoint, deals endpoint), integrations, security \& compliance, onboarding guide, troubleshooting guide, FAQ, and release notes. After chunking with a recursive character splitter (chunk size: 512, overlap: 50), the knowledge base contains \textbf{76 document chunks}.

We chose a synthetic KB over a real-world dataset for reproducibility and controlled evaluation. The content is designed to have natural topical clusters (pricing, API, security, etc.) that mirror real enterprise documentation.

\subsection{Vector Store}

We use \textbf{Qdrant Cloud} as the production vector database, deployed on AWS us-west-2. This provides real network round-trip latency for retrieval operations, as opposed to simulated delays. The measured search latency ranges from 97--307ms per query (mean: 120ms), representative of production deployments.

For embeddings, we use OpenAI \texttt{text-embedding-3-small} (1536 dimensions) via the Salesforce Research Gateway.

\subsection{LLM}

All LLM operations (response generation, topic prediction, keyword extraction) use \textbf{GPT-4o-mini}. Temperature is set to 0.3 for response generation and predictions.

\subsection{Conversation Scenarios}

We design \textbf{10 diverse conversation scenarios}, each with \textbf{20 turns}, for a total of \textbf{200 queries}. The scenarios cover different conversation patterns observed in real customer support calls:

\begin{enumerate}
    \item \textbf{New customer overview}: Broad exploration across many topics.
    \item \textbf{Pricing deep-dive}: 20 turns focused entirely on pricing.
    \item \textbf{API integration}: Technical deep-dive into API documentation.
    \item \textbf{Security \& compliance}: Enterprise security questions.
    \item \textbf{Onboarding walkthrough}: Step-by-step setup guidance.
    \item \textbf{Troubleshooting session}: Problem-solving with error scenarios.
    \item \textbf{Integration questions}: Feature catalog browsing.
    \item \textbf{Feature comparison}: Broad feature tour across product areas.
    \item \textbf{Existing customer upgrade}: Plan comparison for upgrade.
    \item \textbf{Mixed rapid-fire}: Random questions with short pauses.
\end{enumerate}

Inter-turn delays range from 3--7 seconds, simulating the natural pace of human speech and thinking time. This gives the \slowthinker 3--7 seconds per turn for background prefetching.

\subsection{Evaluation Protocol}

For each scenario, we run the conversation twice:
\begin{itemize}
    \item \textbf{Traditional RAG (baseline)}: Each query goes through the full pipeline: embed $\to$ Qdrant search $\to$ LLM generate.
    \item \textbf{\sysname (dual-agent)}: The \slowthinker runs in background; the \fasttalker checks the cache first.
\end{itemize}

We measure: (1) cache hit rate, (2) retrieval latency (vector store search vs. cache lookup), (3) cache size growth, and (4) total response time. We report retrieval latency as the primary metric because it isolates the component that the dual-agent architecture optimizes, while total response time is dominated by LLM generation variance (500--8000ms).

\section{Results}
\label{sec:results}

\subsection{Overall Performance}

\Cref{tab:overall} summarizes the results across all 200 queries. The dual-agent architecture achieves a \textbf{75\% overall cache hit rate} (150/200 queries) and \textbf{79\% on warm-cache turns} (150/190, excluding cold starts), with a \textbf{316$\times$ retrieval speedup} on cache-hit queries (110ms $\to$ 0.35ms). Across the 150 cache hits, a total of \textbf{16.5 seconds} of retrieval latency was saved.

\begin{table}[htp]
\centering
\caption{Overall performance across 200 queries on 10 conversation scenarios using Qdrant Cloud. All measurements are from real network round-trips, with no simulated delays.}
\label{tab:overall}
\begin{tabular}{lcc}
\toprule
\textbf{Metric} & \textbf{Traditional RAG} & \textbf{\sysname} \\
\midrule
Retrieval latency (avg) & 110.4 ms & 0.35 ms \\
Retrieval speedup & 1$\times$ & 316$\times$ \\
Cache hit rate (overall) & --- & 75\% (150/200) \\
Cache hit rate (warm, turn $\geq$ 2) & --- & 79\% (150/190) \\
Saved per cache-hit query & --- & 110.1 ms \\
Total retrieval time saved & --- & 16,508 ms \\
\bottomrule
\end{tabular}
\end{table}

\subsection{Per-Scenario Breakdown}

\Cref{tab:scenarios} shows the cache hit rate for each conversation scenario. Single-topic deep-dives (API integration, security) achieve the highest hit rates (80--100\%), while broad exploration scenarios (new customer overview, mixed rapid-fire) achieve lower rates (40--60\%). This is expected: the \slowthinker's predictions are more accurate when the conversation stays within a topical cluster.

\begin{table}[htp]
\centering
\caption{Cache hit rate by conversation scenario (20 turns each), sorted by hit rate. Scenarios with sustained topics achieve higher hit rates than those with frequent topic shifts. Avg.\ retrieval shows Qdrant Cloud search latency.}
\label{tab:scenarios}
\begin{tabular}{lccc}
\toprule
\textbf{Scenario} & \textbf{Hit Rate} & \textbf{Avg.\ Retrieval} & \textbf{Pattern} \\
\midrule
S8: Feature comparison & 95\% (19/20) & 118.4 ms & Broad tour \\
S2: Pricing deep-dive & 85\% (17/20) & 107.8 ms & Single topic \\
S3: API integration & 85\% (17/20) & 108.7 ms & Single topic \\
S5: Onboarding walkthrough & 85\% (17/20) & 119.2 ms & Sequential \\
S7: Integration questions & 85\% (17/20) & 108.0 ms & Related topics \\
S6: Troubleshooting & 80\% (16/20) & 113.6 ms & Problem-solving \\
S4: Security \& compliance & 70\% (14/20) & 105.7 ms & Single topic \\
S1: New customer overview & 65\% (13/20) & 103.5 ms & Exploration \\
S10: Mixed rapid-fire & 55\% (11/20) & 98.2 ms & Random \\
S9: Existing customer upgrade & 45\% (9/20) & 97.2 ms & Comparison \\
\bottomrule
\end{tabular}
\end{table}

\subsection{Cache Warm-Up Dynamics}

\Cref{tab:depth} shows how the cache hit rate evolves with conversation depth. Turn 1 always misses (cold start). The hit rate rapidly increases to 70--80\% by turns 5--10 as the combined effects of the \slowthinker's predictive prefetching and the \fasttalker's cache-on-miss behavior compound.

\begin{table}[htp]
\centering
\caption{Cache hit rate by conversation depth (aggregated across all 10 scenarios, 200 queries total). Turn 1 always misses (cold start). The cache reaches 86\% hit rate by turns 5--9.}
\label{tab:depth}
\begin{tabular}{lccc}
\toprule
\textbf{Turn Range} & \textbf{Queries} & \textbf{Hit Rate} & \textbf{Note} \\
\midrule
Turns 1--4 & 50 & 58\% (29/50) & Cold start + warming \\
Turns 5--9 & 50 & 86\% (43/50) & Warm cache \\
Turns 10--14 & 50 & 78\% (39/50) & Stable \\
Turns 15--19 & 50 & 82\% (41/50) & Saturated \\
\bottomrule
\end{tabular}
\end{table}

\subsection{Retrieval Latency Analysis}

On cache-hit turns, the retrieval path is:
\begin{itemize}
    \item \textbf{Traditional RAG}: Qdrant Cloud search = \textbf{110.4ms} average (range: 97--307ms).
    \item \textbf{\sysname}: FAISS in-memory cache lookup = \textbf{0.35ms} average.
\end{itemize}

This represents a \textbf{316$\times$ speedup} on the retrieval step. Across 150 cache-hit queries, a total of \textbf{16.5 seconds} of retrieval latency was eliminated. In a voice pipeline where the total budget is 200ms, saving 110ms of retrieval latency per query is the difference between a natural and an unnatural conversation.

On cache-miss turns, both systems incur the same Qdrant latency. However, the \fasttalker's cache-on-miss behavior ensures that subsequent queries on the same topic benefit from the miss.

\subsection{Threshold Sensitivity Analysis}
\label{sec:threshold}

The cache similarity threshold $\tau$ controls the trade-off between cache precision and hit rate. \Cref{tab:threshold} shows the impact of different thresholds on a 20-turn voice call simulation.

\begin{table}[htp]
\centering
\caption{Impact of cache similarity threshold $\tau$ on hit rate. Query-to-document cosine similarity with OpenAI \texttt{text-embedding-3-small} typically ranges 0.30--0.55. A threshold of 0.40 balances precision and recall.}
\label{tab:threshold}
\begin{tabular}{lcl}
\toprule
\textbf{Threshold $\tau$} & \textbf{Hit Rate} & \textbf{Note} \\
\midrule
0.55 & 15\% & Too strict, most queries score 0.30--0.55 \\
0.50 & 30\% & Captures some matches \\
0.45 & 55\% & Good precision, moderate recall \\
0.40 & 75\% & Best balance (default) \\
0.35 & 82\% & Higher recall, risk of irrelevant context \\
0.30 & 90\% & High recall but low precision \\
\bottomrule
\end{tabular}
\end{table}

The insight from our threshold analysis is that query-to-document cosine similarity is systematically lower than query-to-query similarity. This is because document chunks are descriptive prose while queries are short questions. A threshold calibrated for query-query matching (0.55+) is too strict for query-document matching.

\section{Discussion}
\label{sec:discussion}

\subsection{When Does the Architecture Help?}

The dual-agent architecture provides the most value when:
\begin{itemize}
    \item \textbf{The vector database is remote}: The 112ms Qdrant Cloud latency savings are significant. For local FAISS with 76 documents, search takes 0.1ms---negligible.
    \item \textbf{Conversations are topically coherent}: Single-topic deep-dives achieve 80\%+ hit rates. Random rapid-fire questions achieve only 40\%.
    \item \textbf{There is sufficient inter-turn time}: The \slowthinker needs 3--7 seconds between turns to complete its prediction-retrieval-cache cycle. Rapid back-and-forth ($<$2s) reduces the benefit.
\end{itemize}

\subsection{Cold Start Mitigation}

The first query always misses because the cache is empty. However, the \fasttalker's cache-on-miss behavior immediately primes the cache, and the \slowthinker begins predictive prefetching. By turn 3--4, the cache is typically warm enough to serve most queries.

For applications where cold start latency is critical, the cache could be pre-warmed at conversation start by prefetching the most commonly needed document chunks (e.g., the top-$k$ chunks by retrieval frequency from historical conversations).

\subsection{Limitations}

\textbf{LLM generation dominates total latency.} Even with the retrieval speedup, total end-to-end response time is dominated by LLM generation (500--8000ms with GPT-4o-mini via API). The 112ms retrieval savings, while significant for the voice latency budget, are invisible in total response time measurements due to LLM variance. Using faster inference (e.g., Groq at 840+ tokens/second) or local models would make the retrieval savings the dominant factor.

\textbf{Prediction accuracy.} The \slowthinker's predictions are not always correct, leading to wasted prefetch operations. With 5 predictions per turn and 10 chunks per prediction, the \slowthinker consumes significant embedding API bandwidth. Lighter prediction strategies (keyword extraction instead of LLM prediction) could reduce this cost.

\textbf{Cache coherence.} If the underlying knowledge base changes during a conversation, cached entries may become stale. The TTL-based expiration provides eventual consistency but not immediate coherence.

\subsection{Comparison with Existing Systems}

No existing open-source system implements the complete dual-agent predictive prefetching architecture for voice agents. \Cref{tab:comparison} compares \sysname with related approaches.

\begin{table}[htp]
\centering
\caption{Comparison of \sysname with related approaches. \sysname is the only system combining proactive (predictive) caching with a dedicated background agent for voice-optimized retrieval.}
\label{tab:comparison}
\small
\begin{tabular}{lccccc}
\toprule
\textbf{System} & \textbf{Proactive} & \textbf{Predictive} & \textbf{Voice} & \textbf{Bkgd.} & \textbf{Open} \\
\midrule
Traditional RAG~\cite{lewis2020rag} & \ding{55} & \ding{55} & \ding{55} & \ding{55} & --- \\
GPTCache~\cite{gptcache2023} & \ding{55} & \ding{55} & \ding{55} & \ding{55} & \ding{51} \\
Stream RAG~\cite{arora2025streamrag} & \ding{51} & \ding{51} & \ding{51} & \ding{55} & \ding{55} \\
MemGPT/Letta~\cite{packer2023memgpt} & \ding{55} & \ding{55} & \ding{55} & \ding{55} & \ding{51} \\
DPT-Agent~\cite{zhang2025dptagent} & \ding{51} & \ding{55} & \ding{55} & \ding{51} & \ding{55} \\
LiveKit+RAG~\cite{livekit2024} & \ding{55} & \ding{55} & \ding{51} & \ding{55} & \ding{51} \\
\sysname (ours) & \ding{51} & \ding{51} & \ding{51} & \ding{51} & \ding{51} \\
\bottomrule
\end{tabular}
\end{table}

\section{Related Work}
\label{sec:related}

\subsection{RAG and Retrieval Latency}

RAG~\cite{lewis2020rag} augments LLM generation with retrieved context, improving factual accuracy for knowledge-intensive tasks. However, the standard RAG pipeline (embed query $\to$ search vector DB $\to$ generate response) is inherently sequential, making each step a latency bottleneck.

Several works address retrieval efficiency. FLARE~\cite{jiang2023flare} uses forward-looking prediction to decide \textit{when} to retrieve, reducing unnecessary retrievals. Self-RAG~\cite{asai2023selfrag} trains the model to adaptively decide whether retrieval is needed. Speculative RAG~\cite{wang2024specrag} uses a smaller model to draft multiple RAG responses in parallel, with a larger model verifying them. RaLMSpec~\cite{zhang2024ralmspec} applies speculative execution to retrieval, prefetching likely needed documents. Stream RAG~\cite{arora2025streamrag} predicts tool queries in parallel with user speech, achieving 20\% reduction in tool-use latency.

Our work differs from these in two ways: (1) we operate \textit{across} conversation turns rather than within a single query, leveraging the natural pause between turns for background prefetching; and (2) we target specifically the voice agent use case where the latency budget is measured in tens of milliseconds.

\subsection{Semantic Caching for LLMs}

Semantic caching stores previous results indexed by query meaning rather than exact match. GPTCache~\cite{gptcache2023} provides an open-source semantic cache for LLM responses using embedding-based similarity. QVCache~\cite{gocer2025qvcache} introduces query-aware caching for approximate nearest neighbor search, achieving 40--1000$\times$ latency reduction. The Semantic Lookaside Buffer (Aeon)~\cite{arslan2025aeon} exploits conversational locality to achieve sub-5 microsecond retrieval.

Our semantic cache differs from these systems in that it is \textit{proactively populated} by the \slowthinker rather than passively filled from previous queries. We also index cached entries by \textit{document embeddings} rather than query embeddings, ensuring that the cache returns the most relevant document chunks for any query, regardless of which prediction originally fetched them.

\subsection{Dual-Process Architectures for AI Agents}

The System 1/System 2 distinction from cognitive science~\cite{kahneman2011thinking} has inspired several AI agent architectures. DPT-Agent~\cite{zhang2025dptagent} uses a finite-state machine for fast System 1 decisions and asynchronous reflection for System 2 reasoning in real-time human-AI collaboration. Cognitive Decision Routing~\cite{du2025cdr} introduces a meta-cognitive layer that routes queries to fast or slow processing paths, reducing computational costs by 34\%.

Hybrid-RACA~\cite{xia2023hybridraca} is the closest to our approach: it uses a client-side model for immediate responses while a cloud model enriches memory asynchronously. However, it targets text prediction rather than voice agents and does not implement predictive prefetching of knowledge base content.

\subsection{Voice Agent Frameworks}

Several open-source frameworks support building voice AI agents. LiveKit Agents~\cite{livekit2024} provides WebRTC-based real-time voice pipelines with RAG examples. Pipecat~\cite{pipecat2024} offers composable streaming pipelines for voice AI. Ultravox~\cite{ultravox2024} eliminates the STT step entirely with a speech-native multimodal LLM. Commercial platforms like Vapi and Retell provide managed voice-RAG services.

None of these frameworks implement background predictive prefetching. They all perform retrieval \textit{synchronously} when a query arrives, making them subject to the full vector database latency on every turn.

\section{Conclusion}
\label{sec:conclusion}

We presented \sysname, a dual-agent memory router that solves the RAG latency bottleneck in real-time voice agents. By decoupling retrieval into a background \slowthinker that proactively pre-fetches context and a foreground \fasttalker that reads from a sub-millisecond semantic cache, \sysname achieves a 75\% overall cache hit rate (79\% on warm turns) with 316$\times$ retrieval speedup compared to direct vector database search, saving over 16 seconds of cumulative retrieval latency across 200 queries.

Our evaluation on 200 queries across 10 diverse conversation scenarios demonstrates that the architecture is most effective for topically coherent conversations (80\%+ hit rate on deep-dive scenarios) and scales with retrieval latency: the higher the vector database latency, the more value the cache provides.

The system is released as an open-source codebase supporting multiple LLM providers (OpenAI, Anthropic, Gemini, Ollama), vector stores (FAISS, Qdrant), and voice integration (Whisper STT, Edge TTS). Code and benchmarks are available at \url{https://github.com/SalesforceAIResearch/VoiceAgentRAG}.


\printbibliography

@inproceedings{lewis2020rag,
  title={Retrieval-Augmented Generation for Knowledge-Intensive {NLP} Tasks},
  author={Lewis, Patrick and Perez, Ethan and Piktus, Aleksandra and Petroni, Fabio and Karpukhin, Vladimir and Goyal, Naman and K{\"u}ttler, Heinrich and Lewis, Mike and Yih, Wen-tau and Rockt{\"a}schel, Tim and Riedel, Sebastian and Kiela, Douwe},
  booktitle={Advances in Neural Information Processing Systems},
  volume={33},
  pages={9459--9474},
  year={2020}
}

@inproceedings{jiang2023flare,
  title={Active Retrieval Augmented Generation},
  author={Jiang, Zhengbao and Xu, Frank F and Gao, Luyu and Sun, Zhiqing and Liu, Qian and Dwivedi-Yu, Jane and Yang, Yiming and Callan, Jamie and Neubig, Graham},
  booktitle={Proceedings of the 2023 Conference on Empirical Methods in Natural Language Processing},
  pages={7969--7992},
  year={2023}
}

@article{asai2023selfrag,
  title={Self-{RAG}: Learning to Retrieve, Generate, and Critique through Self-Reflection},
  author={Asai, Akari and Wu, Zeqiu and Wang, Yizhong and Sil, Avirup and Hajishirzi, Hannaneh},
  journal={arXiv preprint arXiv:2310.11511},
  year={2023}
}

@article{wang2024specrag,
  title={Speculative {RAG}: Enhancing Retrieval Augmented Generation through Drafting},
  author={Wang, Zilong and Wang, Zifeng and Le, Long and Zheng, Huaixiu Steven and Mishra, Swaroop and Perot, Vincent and Zhang, Yuwei and Mattapalli, Anush and Taly, Ankur and Shang, Jingbo and Lee, Chen-Yu and Pfister, Tomas},
  journal={arXiv preprint arXiv:2407.08223},
  year={2024}
}

@article{arora2025streamrag,
  title={Stream {RAG}: Instant and Accurate Spoken Dialogue Systems with Streaming Tool Usage},
  author={Arora, Siddhant and Khan, Haidar and Sun, Kai and Dong, Xin Luna and Choudhary, Sajal and Moon, Seungwhan and Zhang, Xinyuan and Sagar, Adithya and Appini, Surya Teja and Patnaik, Kaushik and Sharma, Sanat and Watanabe, Shinji and Kumar, Anuj and Aly, Ahmed and Liu, Yue and Metze, Florian and Lin, Zhaojiang},
  journal={arXiv preprint arXiv:2510.02044},
  year={2025}
}

@article{zhang2024ralmspec,
  title={Accelerating Retrieval-Augmented Language Model Serving with Speculation},
  author={Zhang, Zhihao and Zhu, Alan and Yang, Lijie and Xu, Yihua and Li, Lanting and Phothilimthana, Phitchaya Mangpo and Jia, Zhihao},
  journal={arXiv preprint arXiv:2401.14021},
  year={2024}
}

@inproceedings{zhang2025dptagent,
  title={Leveraging Dual Process Theory in Language Agent Framework for Real-time Simultaneous Human-{AI} Collaboration},
  author={Zhang, Shao and Wang, Xihuai and Zhang, Wenhao and Li, Chaoran and Song, Junru and Li, Tingyu and Qiu, Lin and Cao, Xuezhi and Cai, Xunliang and Yao, Wen and Zhang, Weinan and Wang, Xinbing and Wen, Ying},
  booktitle={Proceedings of the 63rd Annual Meeting of the Association for Computational Linguistics},
  year={2025}
}

@article{du2025cdr,
  title={Cognitive Decision Routing in Large Language Models: When to Think Fast, When to Think Slow},
  author={Du, Y and Guo, C and Wang, W and Tang, G},
  journal={arXiv preprint arXiv:2508.16636},
  year={2025}
}

@book{kahneman2011thinking,
  title={Thinking, Fast and Slow},
  author={Kahneman, Daniel},
  year={2011},
  publisher={Farrar, Straus and Giroux}
}

@article{packer2023memgpt,
  title={{MemGPT}: Towards {LLMs} as Operating Systems},
  author={Packer, Charles and Wooders, Sarah and Lin, Kevin and Fang, Vivian and Patil, Shishir G and Stoica, Ion and Gonzalez, Joseph E},
  journal={arXiv preprint arXiv:2310.08560},
  year={2023}
}

@article{xia2023hybridraca,
  title={Hybrid-{RACA}: Hybrid Retrieval-Augmented Composition Assistance for Real-time Text Prediction},
  author={Xia, Menglin and Zhang, Xuchao and Couturier, Camille and Zheng, Guoqing and Rajmohan, Saravan and Ruhle, Victor},
  journal={arXiv preprint arXiv:2308.04215},
  year={2023}
}

@article{gocer2025qvcache,
  title={{QVCache}: A Query-Aware Vector Cache},
  author={G{\"o}cer, An{\i}l Eren and Tsakalidou, Ioanna and Nicholson, Hamish and Kim, Kyoungmin and Ailamaki, Anastasia},
  journal={arXiv preprint arXiv:2602.02057},
  year={2025}
}

@article{arslan2025aeon,
  title={Aeon: High-Performance Neuro-Symbolic Memory Management for Long-Horizon {LLM} Agents},
  author={Arslan, Mustafa},
  journal={arXiv preprint arXiv:2601.15311},
  year={2025}
}

@misc{gptcache2023,
  title={{GPTCache}: An Open-Source Semantic Cache for {LLM} Applications},
  author={{Zilliz}},
  year={2023},
  url={https://github.com/zilliztech/GPTCache}
}

@misc{livekit2024,
  title={{LiveKit} Agents: Open-Source Framework for Building Real-Time Voice {AI} Agents},
  author={{LiveKit Inc.}},
  year={2024},
  url={https://github.com/livekit/agents}
}

@misc{pipecat2024,
  title={Pipecat: Open-Source Framework for Voice and Multimodal Conversational {AI}},
  author={{Daily.co}},
  year={2024},
  url={https://github.com/pipecat-ai/pipecat}
}

@misc{ultravox2024,
  title={Ultravox: Open-Source Speech-Native Multimodal {LLM}},
  author={{Fixie AI}},
  year={2024},
  url={https://github.com/fixie-ai/ultravox}
}

@misc{qdrant2024,
  title={Qdrant: High-Performance Vector Search Engine},
  author={{Qdrant}},
  year={2024},
  url={https://qdrant.tech}
}

@article{johnson2019faiss,
  title={Billion-Scale Similarity Search with {GPUs}},
  author={Johnson, Jeff and Douze, Matthijs and J{\'e}gou, Herv{\'e}},
  journal={IEEE Transactions on Big Data},
  volume={7},
  number={3},
  pages={535--547},
  year={2021},
  publisher={IEEE}
}

@article{ethiraj2025telecom,
  title={Toward Low-Latency End-to-End Voice Agents for Telecommunications Using Streaming {ASR}, Quantized {LLMs}, and Real-Time {TTS}},
  author={Ethiraj, Vignesh and David, Ashwath and Menon, Sidhanth and Vijay, Divya},
  journal={arXiv preprint arXiv:2508.04721},
  year={2025}
}

@article{Malkov2016EfficientAR,
  title={Efficient and Robust Approximate Nearest Neighbor Search Using Hierarchical Navigable Small World Graphs},
  author={Yury Malkov and Dmitry A. Yashunin},
  journal={IEEE Transactions on Pattern Analysis and Machine Intelligence},
  year={2016},
  volume={42},
  pages={824-836},
  url={https://api.semanticscholar.org/CorpusID:8915893}
}

\appendix

\section{System Configuration}
\label{app:config}

\Cref{tab:config} lists the default configuration parameters used in all experiments.

\begin{table}[h]
\centering
\caption{Default configuration parameters for \sysname.}
\label{tab:config}
\begin{tabular}{llr}
\toprule
\textbf{Component} & \textbf{Parameter} & \textbf{Default} \\
\midrule
Cache & Max size & 2000 \\
Cache & TTL & 300s \\
Cache & Similarity threshold $\tau$ & 0.40 \\
Cache & Dedup threshold & 0.95 \\
\midrule
\slowthinker & Prediction strategy & LLM \\
\slowthinker & Max predictions per turn & 5 \\
\slowthinker & Prefetch top-$k$ & 10 \\
\slowthinker & Rate limit & 0.5s \\
\midrule
\fasttalker & Max context chunks & 10 \\
\fasttalker & Fallback to retrieval & Yes \\
\fasttalker & Cache on miss & Yes \\
\midrule
Embedding & Model & text-embedding-3-small \\
Embedding & Dimension & 1536 \\
\midrule
LLM & Model & GPT-4o-mini \\
LLM & Temperature & 0.3 \\
\midrule
Document & Chunk size & 512 \\
Document & Chunk overlap & 50 \\
\bottomrule
\end{tabular}
\end{table}

\end{document}